\documentclass[manuscript,screen,nonacm]{acmart}

\usepackage{svg}

\usepackage{enumitem}
\setlist[enumerate]{itemsep=4pt}
\setlist[description]{labelindent=0.5cm, leftmargin=!,listparindent=\parindent, parsep=\parskip,itemsep=4pt}

\usepackage{verbatim}

\AtBeginDocument{%
  }

\setcopyright{acmlicensed}
\copyrightyear{2026}
\acmYear{2026}
\acmDOI{XXXXXXX.XXXXXXX}

\setcopyright{none}
\renewcommand\footnotetextcopyrightpermission[1]{} 

\acmJournal{JRC}

\begin{document}

\title{From Process to Evidence: How Computing Can Ground Appropriate Reliance on Legal AI}
\author{James Bryan Williams}
\correspondingauthor
\email{jbwilli1@nycourts.gov}
\orcid{0000-0002-3352-4076}
\affiliation{%
  \institution{N.Y. Supreme Court, Appellate Division}
  \city{Rochester}
  \state{New York}
  \country{USA}
}
\affiliation{%
  \institution{University at Buffalo}
  \city{Buffalo}
  \state{New York}
  \country{USA}
}

\renewcommand{\shortauthors}{Williams}

\begin{abstract}
Lawyers and self-represented litigants are already using artificial intelligence (AI) to draft legal documents, and courts are responding with rules. After more than 1,500 cases involving AI hallucinations, lawyers have been instructed to perform careful, independent review of AI-assisted filings. Discharging these duties requires what the human-computer interaction (HCI) literature calls ``appropriate reliance,'' which cannot be calibrated without evidence on how often, how badly, and how detectably these tools fail at legal work. Existing research barely describes any of the three.

We analyze the official record (i.e., the complete set of AI-related rules and guidelines) of the New York court system. These artifacts explicitly ask for evidence that does not exist (e.g., error rates, do-not-use lists). In its place they invoke procedure, including training mandates, checklists, and uncalibrated human review. The burden of compliance falls hardest on those least equipped to bear it: legal aid programs are told to track their own error rates, and judges are left to improvise their own tests.

The paper makes four contributions: (1) a mapping from the legal duties to concepts in HCI; (2) a set of requirements elicited from the official record; (3) an analysis of how the legal system substitutes process for evidence; and (4) a research agenda, including task taxonomies, shared error metrics, maintained benchmarks, and test harnesses for evaluations on private data. The computing community must supply what the justice system lacks; in doing so, it can help close, rather than widen, the justice gap.
\end{abstract}

\ccsdesc[500]{Applied computing~Law}
\ccsdesc[500]{Social and professional topics~Governmental regulations}
\ccsdesc[300]{General and reference~Evaluation}
\ccsdesc[300]{Human-centered computing~Empirical studies in HCI}

\keywords{legal AI, appropriate reliance, generative artificial intelligence,
verification, evaluation, professional responsibility, access to justice,
courts, large language models}

\maketitle

\section{Introduction}
\label{sec:intro}

\textit{Artificial intelligence} (AI) is being adopted across the legal sector faster than its institutions can govern it. In a recent survey, the share of legal professionals using \textit{generative AI} (genAI) nearly doubled in a year, from 14\% to 26\% in 2025 \cite{tri2025},  and 95\% of respondents across professional services expected genAI to be central to their workflow within five years. A subsequent survey reported that organization-wide use reached roughly 40\% in 2026 \cite{tr2026profservices}. GenAI tools are now embedded in the research platforms lawyers use daily \cite{magesh2025}. However, nearly half of law firm respondents reported no AI policy, and most had received no training in genAI use.  A 2026 survey of in-house legal leaders found the same gap on the demand side: 83\% could not measure whether their AI spending was working, and the leading obstacle (for non-adopters) was uncertainty about which tools are useful for legal work \cite{axiom2026}.

AI tools are also being used to address the ``justice gap.'' \textit{Self-represented litigants} (SRLs) already use \textit{large language models} (LLMs) to analyze and draft legal documents; in many cases, the model is the only help available, since low-income Americans receive insufficient help for 92\% of their civil legal problems \cite{lsc2022}. Legal aid clinics are also adopting AI quickly. In a 2024 Illinois survey, 90\% of legal aid professionals reported little or no AI use, but most said their organizations were open to adoption \cite{ltf_ai_survey_2024}; a year later, 74\% reported AI use \cite{everlaw2025}.

Adoption has outpaced lawyers' ability to judge when these tools can be relied on. In \textit{Mata v. Avianca}, lawyers were sanctioned for citing fictitious cases \cite{mata2023} (see also \cite{noland2025}), and the AI Hallucination database now logs over 2293 such incidents \cite{charlotin}. Recent research suggests the scale of the problem: general-purpose LLMs answer case-law questions incorrectly in 58--82\% of queries \cite{dahl2024}, and commercial legal research tools still produce incorrect or unsupported answers in 17--33\% \cite{magesh2025}. Bar associations offer \textit{continuing legal education} (CLE) programs on these risks (e.g., \cite{cle2026,nysba_ai_cle_2026_trust_but_verify}).

\begin{figure}[h!]
    \centering
    \includegraphics[width=0.85\linewidth]{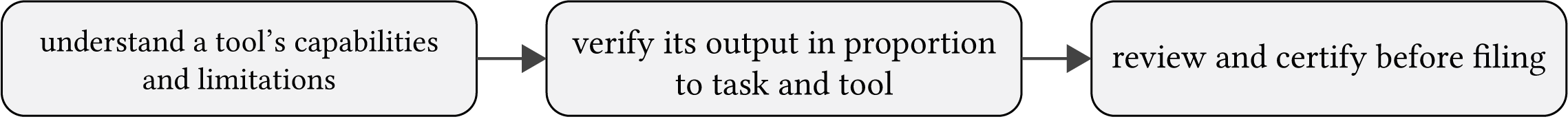}
    \caption{Obligations for the use of Legal AI.}
    \label{fig:trio}
    \Description{The three major duties for the use of legal AI, shown as three boxes in sequence. Understand the tool's capabilities and limitations. Verify its output in proportion to task and tool. Review and certify before filing.}
\end{figure}

Legal institutions are responding by imposing duties and issuing guidelines. In 2024, the \textit{American Bar Association} (ABA) issued \textit{Formal Opinion 512} \cite{aba_formal_512}, mapping the ABA \textit{Model Rules of Professional Conduct} (MRPC) \cite{abarules} onto AI. It advises lawyers to maintain a reasonable understanding of any tool's ``capabilities and limitations'' and to verify its output in proportion to task and tool. In March 2026, the \textit{New York Unified Court System} (NYUCS) adopted 22 NYCRR Part 161 \cite{nys_courtrules_2026}, effective June 1. The policy permits AI use and offers a \textit{model rule} that judges may adopt. 
Where adopted, users are expected to understand each tool's ``capabilities and limitations.'' They are also required to ``carefully review'' AI-assisted filings and ``independently ensure'' they contain no ``fabricated or fictitious cases, statutes, or other material.''
Signatures attest that the review occurred; sanctions may follow if it did not. Signer duties under 22 NYCRR 130-1.1a already bind every attorney and party, including SRLs. The NYUCS also binds judges and court employees \cite{ucspolicy2025}, and bar associations recommend the same safeguards \cite{nysba2026a2j}. Across these instruments, careful human review is treated as sufficient protection. Figure \ref{fig:trio} synthesizes the obligations for users of legal AI.\\

Discharging these duties requires what the \textit{human-computer interaction} (HCI) literature calls ``appropriate reliance'' \cite{leesee2004,schemmer2023}: relying on a tool when it is right, and not relying on it when it is wrong. Risk analysis decomposes failure into three measurable dimensions \cite{stamatis2003fmea,iec60812}: (1) how often a failure occurs; (2) how severe it is; and (3) how likely it is to be caught before it does harm. Reliance cannot be calibrated without evidence on all three, and no such body of evidence exists for legal AI. Existing studies measure failure frequency for particular tools at single points in time, but they do not grade severity or measure detectability, and their methods differ enough that results cannot be compared. The \textit{New York State Bar Association} (NYSBA) has acknowledged that independent studies and benchmarks are not available \cite[p.48]{nysba2026ethics}.

Since it is well-established that AI tools can fail, the rules and guidelines rely on compliance and human review at the point of use. The result is a \textit{misallocation}: responsibility for evaluating AI is passed to those least able to perform it. Rules, guidelines, and ethics opinions instruct practitioners to test AI tools, assess their reliability, or delegate these tasks to experts---but the methods do not exist, and neither does the expert role. At least one judge has already improvised reliability experiments while adjudicating a dispute \cite{weber2024}. Legal aid clinics and trial judges should not be measuring AI reliability; instead, the legal system needs both a body of evidence and subject matter experts who can apply it. The computing community is uniquely positioned to provide both.

This article makes four contributions:
\begin{enumerate}
    \item We map the legal duties onto human factors and HCI research. The rules speak of careful review, independent judgment, and verification proportional to the task; we show that these phrases reduce to a single construct, \textit{appropriate reliance} on an imperfect tool. This concept has been studied extensively, with a rich literature of measures and experimental designs \cite{leesee2004,schemmer2023}. The missing element is \textit{legal conditioning}: a description of AI failure on legal tasks.
    \item We provide a  analysis for computing research. We analyze the official record of the New York court system (i.e., court rules, judicial policies, bar association reports, and guidelines). It repeatedly calls for evidence that does not exist, including error rates, ``do-not-use'' determinations, and independent testing.
    \item We show that legal institutions consistently make a category error. When an institution cannot find evidence or expertise, it relies on \textit{process}: training mandates, checklists, review duties, and approval workflows. 
    \item We develop a research agenda for the computing community, including task taxonomies, shared error metrics, maintained benchmarks, and test harnesses that legal institutions can use to evaluate AI tools on private datasets.
\end{enumerate}

Other fields have recognized that AI tools require an evidence system (e.g., \cite{Khlaaf2023Toward}). ``FinTech'' scholarship synthesizes standards and legal instruments into concrete evidence controls, framing compliance as ``an evidence system design problem rather than a model performance challenge''~\cite{staley2026}.  Computing practitioners and state governments studying AI governance gaps have similarly noted that audit and incident-recording infrastructure is largely absent (e.g.,~\cite{jimenezgomez2025government,nyag2024symposium}).

The article proceeds as follows. Section~\ref{sec:record} introduces the official record of the NY court system. Section~\ref{sec:reliance} maps legal duties to appropriate reliance. Section~\ref{sec:requirements} elicits requirements, and Section~\ref{sec:substitution} discusses the substitution of process for evidence. Section~\ref{sec:evidence} measures existing evidence against the requirements. Section~\ref{sec:agenda} sets out a research agenda, while Section~\ref{sec:conclusion} concludes the paper with some suggestions for future work.

\section{The Governance Record}
\label{sec:record}

This paper studies governance documents (the ``official record'') from the NYUCS. New York has more resident attorneys than any other US state (193,536 as of 2025 \cite{nycourts2025annual}). In June 2026 the NYUCS adopted a statewide policy and optional model rule governing the use of AI in filings~\cite{nys_courtrules_2026}. There is a complete documentary trail leading up to that rule, including public hearings, bar association reports, ethics guidelines, and training materials. Focusing on one state keeps the analysis tractable.

The corpus selection rule is simple: a document is included if it binds or guides actors in the New York justice system, directly or by reference. National documents are considered only where New York's own documents incorporate them, as with the ABA's ethics opinion~\cite{aba_formal_512} and a proposed federal evidence rule~\cite{evidence_advisory_report_2025}. Table~\ref{tab:corpus} lists the documents. The corpus covers competence and verification duties only; other duties (e.g., confidentiality, fees, advertising) are out of scope.

\subsection{Practitioners}

Three instruments govern the use of AI by New York lawyers:

\begin{enumerate}
\item \textbf{The national ethics baseline.}

In 2024, the ABA issued Formal Opinion 512 on genAI~\cite{aba_formal_512}. Lawyers must (i) maintain a reasonable understanding of a tool's ``capabilities and limitations,'' and (ii) verify outputs, with effort proportional to the task. They may ``draw on the expertise of others'' to meet these duties.

\item \textbf{The duty of competence.}

Rule 1.1 of the \textit{New York Rules of Professional Conduct} (NYRPC)~\cite{nyrpc} defines ``competence'' in representation. Lawyers must possess the knowledge and skill ``reasonably necessary for each matter.'' Comment 8 extends this duty to tools: lawyers should ``keep abreast of the benefits and risks associated with technology the lawyer uses.''

\item \textbf{Court rules.}

22 NYCRR Part 161~\cite{nys_courtrules_2026} applies to every attorney and party (including SRLs) in the NYUCS. It contains a model rule for the use of AI in criminal and civil cases. Users must understand each tool's ``capabilities and limitations,'' ``carefully review'' AI-assisted filings, and ``independently ensure'' they contain no ``fabricated or fictitious cases, statutes, or other material.'' A signature certifies the review; sanctions may follow if the requirements are not met.
\end{enumerate}

\noindent
New York's bar associations have produced numerous artifacts concerning the use of AI by lawyers:

\begin{enumerate}

\item \textbf{The task force report.}

The NYSBA's \textit{Task Force on Artificial Intelligence} delivered the first comprehensive treatment in 2024~\cite{nysba2024taskforce}. It proposed expanding the competence duties to cover how AI tools operate, and raised the possibility of a \emph{duty to use} AI, which would make the evidence problem unavoidable (see also \cite{nacdl_parity_2026}).

\item \textbf{The access-to-justice hearing.}

In 2025, a NYSBA committee held a public hearing on AI and access to justice; its 2026 report collects written testimony from judges, court administrators, legal aid executives, technology vendors, and funders~\cite{nysba2026a2j}.

\item \textbf{The ethics guidelines.}

The NYSBA's \textit{Committee on Artificial Intelligence and Emerging Technologies} (CAIET) issued ethics guidance in 2026, applying the NYRPC to AI~\cite{nysba2026ethics}. CAIET reviewed the empirical studies, questioned whether ``supervision-without-understanding'' can work, then prescribed vetting for ``reliability, accuracy, and security'' anyway.

\item \textbf{The city bar's opinion.}

\textit{New York City Bar Association} (NYCBA) Formal Opinion 2024-5 addressed genAI across the conduct rules~\cite{nycbar2024}, advising lawyers to supplement AI-generated research with human research and warning that guidance must stay general because the tools evolve quickly.

\item \textbf{The training materials.}

NYSBA CLE programs are teaching the profession what these duties mean in practice~\cite{cle2026}; the \textit{New York State Academy of Trial Lawyers} is actively producing CLE programs on AI (e.g.,~\cite{nysatl2025cle}).

\end{enumerate}
These instruments combine a concrete step with an unknown. Citation checking is a defined task with adequate tool support (e.g., Westlaw Edge Quickcheck). Understanding the capabilities and limitations of genAI tools is another matter: lawyers are obligated to draw on a body of evidence that does not exist.

\subsection{Self-Represented Litigants}

Signer duties under 22 NYCRR 130-1.1a bind all parties to an action, not just attorneys. Where a judge adopts Part 161's model rule~\cite{nys_courtrules_2026}, an SRL who uses AI faces the same review and certification expectations as an attorney. Filing certifies the review, and sanctions may follow if it is unsatisfactory. Thus, the burden falls on the parties least able to access the justice system (e.g., \cite{julien2026arthur}).


AI use in court filings is no longer anecdotal: in a sample of 1,600 federal civil complaints, detector-flagged AI-generated text rose from roughly 1\% in 2023 to about 18\% in 2026~\cite{shahlevy2026}. For SRLs, AI is double-edged rather than simply burdensome. Although AI can make an otherwise incoherent filing legible~\cite{shahlevy2026}, it can also introduce fabricated or overstated authority.\footnote{Some judges report that AI is improving the organization of SRL filings without improving SRL arguments~\cite{canadianlawyer2026genai}.} There does not appear to be any research on whether SRLs can assess the quality of AI-generated legal writing (Section~\ref{sec:agenda}).

\subsection{Court Employees}


The NYUCS \textit{Interim AI Policy} (IAIP)~\cite{ucspolicy2025} governs every judge and court employee. The IAIP is the work of the \textit{Advisory Committee on Artificial Intelligence and the Courts} (ACAIC), created in April 2024~\cite{advisorycommittee2024}. The ACAIC is an advisory body, not a research institute; its members are judges, court administrators, practicing attorneys, and academics, serving intermittently as their regular duties allow. The structure is suited to issuing guidance, not producing evidence. 

Appendix 9 of the ACAIC's annual report to the judiciary ~\cite{nyucs2025advisory} provides ethical considerations and recommendations for judicial use of AI. The committee reconciled the IAIP with 22 NYCRR Part 100, stating that a judge may delegate drafting, but not judgment, to AI. Judges who use AI tools should independently validate AI output before relying on it. The recommendations are generally consistent with the duties that Part 161 implies for attorneys. (The advice is timely, as judicial adoption of AI is also significant; in a recent survey, more than 60\% of 112 responding federal judges report using at least one AI tool, with 30\% using AI for legal research \cite{jaitley2026federal}.)

The IAIP also grants the UCS \textit{Division of Technology and Court Research} (DoTCR) control over AI tools. In general these fall into two categories:
\begin{enumerate}
\item General-purpose AI tools, ``whether operating on a public model or on a private model,'' are deemed not suitable ``for legal writing and legal research, as they may produce incorrect or fabricated citations and analysis.''
\item Ancillary AI tools, like those used for translation and transcription, are deemed less sensitive than those used for substantive legal tasks (e.g., drafting).
\end{enumerate}
DoTCR approval of an AI tool signifies that the product is technologically safe to use, but not that it is suitable for any particular task. Finally, the IAIP requires mandatory AI training for the entire workforce.

\begin{table*}[t]
\caption{The corpus: every document analyzed in this paper. National instruments annotated with $\dagger$ are included because New York's own documents incorporate them by reference.}
\label{tab:corpus}
\small
\begin{tabular}{@{}p{0.26\linewidth}p{0.17\linewidth}cp{0.16\linewidth}p{0.25\linewidth}@{}}
\toprule
\textbf{Document} & \textbf{Issued by} & \textbf{Year} & \textbf{Reach} & \textbf{Role in this paper} \\
\midrule
NYRPC Rule 1.1 and Comment 8 & NYUCS (rule); NYSBA (comment) & 2012/2018 & NY attorneys & Competence duty; ``benefits and risks'' (\S\ref{sec:record}) \\
ABA Formal Opinion 512$^\dagger$ & ABA & 2024 & National baseline & Understanding and proportional verification; ``expertise of others'' (\S\ref{sec:record}, \S\ref{sec:requirements}) \\
22 NYCRR Part 161 & NYUCS & 2026 & NY attorneys and parties & Review and certification duty (\S\ref{sec:record}, \S\ref{sec:requirements}) \\
Interim AI Policy (IAIP) & NYUCS & 2025 & NY judges and court staff & ``Not suitable'' exclusion; approval vs.\ suitability (\S\ref{sec:record}, \S\ref{sec:substitution}) \\
ACAIC Annual Report & NYUCS ACAIC & 2025 & NY judges and court staff &
  committee's limitation to guidance
  (\S\ref{sec:record}, \S\ref{sec:substitution}); guidelines for judicial use of AI \\
Task Force on AI report & NYSBA & 2024 & NY attorneys & First comprehensive treatment; the emerging duty \emph{to use} (\S\ref{sec:requirements}) \\
AI and Access to Justice report & NYSBA & 2026 & NY attorneys & The demand specification: error rates, independent testing (\S\ref{sec:requirements}) \\
Ethics Guidelines & NYSBA CAIET & 2026 & NY attorneys & Vetting duty; benchmark concession (\S\ref{sec:requirements}, \S\ref{sec:substitution}) \\
Formal Opinion 2024-5 & NYCBA & 2024 & NY attorneys & Redundancy rule; security-only experts (\S\ref{sec:reliance}--\ref{sec:substitution}) \\
CLE materials & Various providers & 2025--26 & NY attorneys & What the profession is taught (\S\ref{sec:substitution}) \\
\bottomrule
\end{tabular}
\end{table*}

\subsection{The Official Record and its Implications}

Table~\ref{tab:corpus} lists the official record of the NYUCS. Taken as a whole, it converges on a sequence of instructions: (1) understand AI tools; (2) verify their outputs; and (3) review and certify before filing (see Figure \ref{fig:trio}). The sequence is consistent across document types (e.g., binding rules, institutional policies) and across organizations. However, the rules and guidelines do not suggest how these instructions are to be obeyed. Section~\ref{sec:reliance} maps them to appropriate reliance, while Sections~\ref{sec:requirements} and~\ref{sec:substitution} argue that the justice system lacks a sufficient body of knowledge.

\newpage
\section{From Duties to Appropriate Reliance}
\label{sec:reliance}

The duties and guidance converge on a three-part structure: (1) understand a tool's limits; (2) verify its output in proportion to the task and tool; and (3) review and certify before filing. Each part has a formal counterpart in research on human use of automation, developed in human factors and extended to AI in HCI~\cite{leesee2004,schemmer2023,bo2025,Kim2025}.

\subsection{Trust and Reliance}

Following the human factors literature, \textit{trust} is an attitude toward a tool, while \textit{reliance} is the behavior of acting on its output~\cite{leesee2004}. Trust does not determine reliance: combined with other attitudes (e.g., perceived risk, self-confidence) it can form an intention to rely, which becomes reliance behavior if the right environmental and cognitive factors are present. Trust guides reliance when ``complexity and unanticipated situations'' make complete understanding impractical~\cite{leesee2004}.

Trust is not foundational for the legal duties. The official record does not emphasize it: trust appears briefly as a goal to support or an asset to protect (e.g.,~\cite{nysba2026a2j}), but the documents that impose duties do not invoke it (e.g.,~\cite{aba_formal_512,ucspolicy2025,nys_courtrules_2026}). This is understandable, since law regulates conduct, not attitudes. The duties instead obligate lawyers to verify a tool's output in proportion to task and tool. Opinion 512~\cite{aba_formal_512} states that the ``appropriate amount of independent verification or review\ldots will necessarily depend on the GAI tool and the specific task that it performs.'' A lawyer may perform less independent verification or review when prior testing gives a ``reasonable basis for relying on its results.'' The standard speaks of verification and reliance, not trust.

\newpage
\subsection{Appropriate Reliance}
\label{sec:appropriate_reliance}

\begin{figure}[t!]
    \centering
    \includegraphics[width=0.9\linewidth]{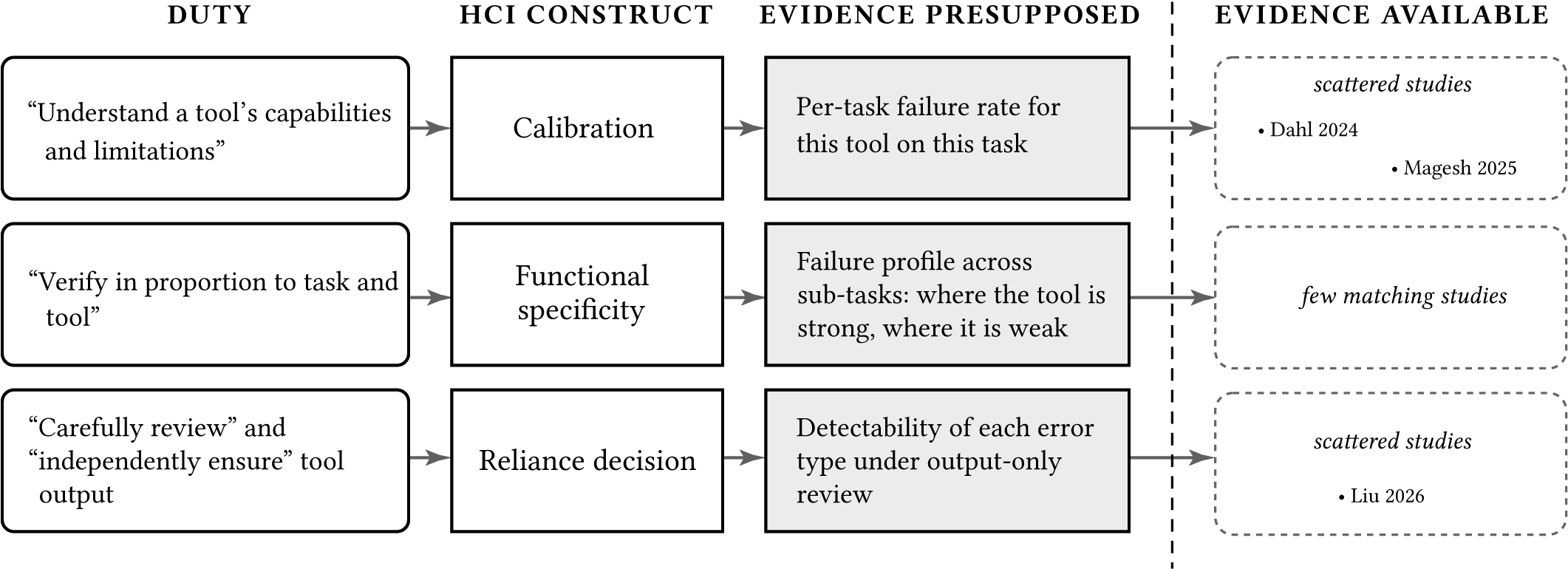}
    \caption{The Legal Duties mapped to Evidence.}
    \label{fig:fig1_duty_construct_evidence}
    \Description{A diagram of the legal duties imposed by the state of New York upon parties and how they map to HCI constructs, presupposed evidence, and available evidence. Duties: (1) understanding a tool's capabilities and limitations; (2) verify in proportion to task and tool; and (3) carefully review and independently ensure tool output.}
\end{figure}

Each legal duty has a counterpart in human factors (see Figure \ref{fig:fig1_duty_construct_evidence}):
\begin{enumerate}

\item Understanding a tool's limits is \textit{calibration}: ``the correspondence between a person's trust in the automation and the automation's capabilities''~\cite{leesee2004}. We use calibration for the epistemic correspondence itself, setting aside the trust that mediates it: a lawyer is well calibrated when confidence in a tool matches what it can actually do.

\item Verifying in proportion to task and tool is \textit{functional specificity}: trust that reflects the ``capabilities of specific subfunctions and modes'' rather than the system as a whole~\cite{leesee2004}. Proportional verification presupposes knowing where a tool is strong and where it is weak, task by task.

\item Reviewing before relying is the \textit{reliance decision}: accept the output when it is right, override it when it is wrong~\cite{schemmer2023}.

\end{enumerate}

Together, these constitute \emph{appropriate reliance}: output by output, adopting what is correct and rejecting what is incorrect~\cite{schemmer2023}. It is measurable. For a given tool and task, measurement involves two rates: how often correct output is accepted, and how often incorrect output is caught. Decades of work in cockpits, control rooms, and clinics have measured these quantities~\cite{leesee2004,parasuraman1997}, and a recent synthesis covers more than fifty studies of appropriate reliance on genAI~\cite{passi2024}.

\subsection{Human Review and its Limits}
\label{sec:human_review_limits}

The second rate  (i.e., how often incorrect output is caught) is the most critical. AI-assisted legal drafting produces two kinds of error. 
\begin{enumerate}
    \item Errors of \textit{commission} are present in the output: a fabricated citation (e.g.,~\cite{noland2025}), a misquoted source, or a misstated holding (e.g., \cite{landberg2026}). These can be checked against the record (e.g., Westlaw).
    \item Errors of \textit{omission} are not: a controlling authority that was missed, or a material fact left out. Nothing in the output points to what is absent; the only way to find it is to redo the work the tool was meant to perform. Review of the output cannot discharge the duty, even in principle. For example, a New York judge identified ``jurisdictional slippage'' (e.g.,  ignoring controlling Appellate Division authority) as a major problem with AI-generated filings \cite{lebovits2026courts}.
\end{enumerate}


The errors made by genAI tools are not the same as those made in \textit{technology-assisted review} (TAR)~\cite{grossman_cormack_2011}. TAR determines whether a fixed, finite set of documents is responsive: reliability is measured by \textit{recall} (the share of responsive documents found) and \textit{precision} (the share of selections that are responsive). The population is closed and the denominator is known. A generated document, by contrast, is a particular writing, not a sample from a space of writings. Statistical methods do not apply, and a judgment about a particular output does not generalize.

\subsection{What do the Legal Duties Entail?}
\label{sec:legal-duties-entail}

The first element is a form of calibration: understanding the capabilities and limits of an AI tool. Calibration is task-specific: evidence about AI tools in general, or on other tasks, is not sufficient. The lawyer needs evidence that a particular tool, applied to a particular task, produces outputs that can be relied upon.

The second element asks for verification in proportion to task and tool (i.e., functional specificity). Verification substitutes for trust: a lawyer, unlike an airline pilot, can examine the output before relying on it. If she can verify independently (e.g., by checking for hallucinated citations), she is obligated to do so. If verification is costly, she has several options:

\begin{enumerate}

\item Decline the task.

\item Forego use of the tool.

\item Decompose the task and use the tool only on subtasks that are calibrated and verifiable.

\item Use the tool with full verification, even at high cost (e.g., by redoing the work).

\item Use the tool with partial verification: verify what can be verified, accept residual risk on the rest, and certify.

\end{enumerate}

The duties permit these options, though some may still expose the lawyer to liability. What she cannot do is rely on the tool without verification or review.

The third element concerns review and certification. Review is fundamental to the rules of professional conduct, while certification is an accountability mechanism. Certifying a filing does not assert that the lawyer's reliance was appropriate, only that she takes personal responsibility for the filing, whatever the tool produced.

\subsection{The Missing Body of Evidence}

Consider a lawyer who uses genAI to draft a filing. Where a judge has adopted Part 161's model rule, careful review is required~\cite{nys_courtrules_2026}, so she must allocate effort: what to check, what to keep, what to redo. Legal writing involves numerous tasks (e.g., checking whether a quotation is faithful, whether an argument overstates its authority, or whether the analysis is sound). Each costs time, and she has no evidence about where this tool, on this task, tends to fail (see \cite[p.232]{magesh2025}). The duty of review may be discharged in form only: certification indicates that a review occurred, but whether it was \emph{effective} depends on unmeasured failure rates.

Now suppose the same lawyer operates under Opinion 512, which permits less verification when prior testing gives a ``reasonable basis for relying on its results''~\cite{aba_formal_512}. She has used the tool many times to summarize short documents and now applies it to draft arguments from a larger set of lengthy ones; she then uses a tool fine-tuned for the law of state X to draft filings in state Y (see~\cite{withers2026}). The tasks differ in ways she may not see, and a tool that performs well on one may fail on the other.

Appropriate reliance requires measurements of error rates that vary by task and error type. A misspelled word, a fabricated citation, a missing authority, and a misinterpretation of an argument have very different detection profiles. The official record leaves evidence-gathering to lawyers, using whatever data is at hand. A lawyer who cannot run such a test has no basis to verify less, and so must either verify everything or assume significant risk (i.e., the ``verification-value paradox''~\cite{yuvaraj2025}; see \cite[p.232]{magesh2025}).\footnote{\ The paradox is that AI efficiency gains  provide greater impetus to manually verify outputs, which often leaves the net value negligible.} In addition, prior experience is not always a useful guide for technical reasons: AI models change continuously, and performance is often not stable within a named version \cite{Chen2024How}.

No external body of evidence exists for any of these rates. One recent study~\cite{magesh2025}, cited by Opinion 512~\cite{aba_formal_512}, found that genAI tools are difficult to assess for trust or safety and that their documentation does not state which use cases warrant caution. Quantitative information about risks and benefits is generally unavailable~\cite[p.48]{nysba2026ethics}. Legal educators advise lawyers to ``trust but verify''~\cite{nysba_ai_cle_2026_trust_but_verify} tools whose reliability has not been measured on similar tasks.\footnote{As another example, a bar association recommends tailoring AI chatbot supervision to the ``likelihood that ethical problems may arise''~\cite{nycbar2024}.}

The remainder of this paper considers two questions: what evidence does appropriate reliance require, and can the justice system gather it? Sections~\ref{sec:requirements} and~\ref{sec:substitution} show that the official record calls for strong evidence and, failing to find it, substitutes process. Section~\ref{sec:evidence} shows that the current research literature is insufficient.

\section{The Implicit Requirements for Legal AI}
\label{sec:requirements}

Section~\ref{sec:reliance} argued that the evidentiary basis for appropriate reliance is missing. This section begins to fill in the gap. The underlying question is simple: what must be built to ground appropriate reliance on legal AI? How can error rates be measured, and how can calibration, functional specificity, and reliance decisions (see Section~\ref{sec:appropriate_reliance}) be understood in this domain?
We address these questions through \textit{requirements analysis}, treating the official record as elicitation material. Requirements engineering distinguishes between stakeholder voices, formal constraints, and unresolved questions; the official record contains all three:

\begin{itemize}

\item The public hearings~\cite{nysba2026a2j} are a set of stakeholder position statements.

\item The rules and policies describe constraints and goals.

\item The training materials raise open questions and expose gaps in current guidance.

\end{itemize}

\textit{Document analysis} is an established elicitation technique when stakeholder statements are on record~\cite{bowen2009}. The target is an \textit{evaluation infrastructure} for legal AI: shared, continuously maintained benchmarks; task taxonomies; and measurement protocols that characterize how often and how detectably a tool fails at a given legal task. The infrastructure provides the per-task evidence that lets verification be calibrated instead of guessed.

\subsection{Stakeholders}

The stakeholders are easily identified from the official record:
\begin{itemize}

\item Courts must decide what tools to approve and what filings to accept~\cite{ucspolicy2025,nys_courtrules_2026}.

\item Attorneys must calibrate verification to tool and task~\cite{aba_formal_512}.

\item  SRLs bear some of the same signer duties as attorneys (e.g., review and certification) with no training~\cite{nys_courtrules_2026}.

\item \textit{Legal services organizations} (LSOs) must choose tools under resource constraints and answer to funders~\cite{nysba2026a2j}.

\item Vendors seek adoption.

\end{itemize}

Their interests often conflict. Vendors have a strong incentive to encourage adoption even when evidence is scarce; they ask whether it is ``ethical not to use'' the tools~\cite[p.64]{nysba2026a2j}; other stakeholders warn courts not to take vendor claims ``at face value''~\cite[p.61]{nysba2026a2j}.

The burden of verification does not fall equally. A recent study found that risks concentrate on the litigants least able to absorb them: those in lower courts or less prominent jurisdictions, those seeking more complex legal information, and those least able to gauge how far to trust the output~\cite{dahl2024}. SRLs use AI, yet are the least capable of verifying its output.

\subsection{Functional Requirements}

The functional requirements state what the evaluation system must do. We divide them into two categories: outputs and mechanisms.\\

\noindent
\textbf{Functional Requirements: Outputs}

\noindent
These requirements describe what the system must produce:

\begin{description}

    \item[Report failure rates by task.] The system must provide, by measurement rather than assertion, the per-task rate at which a tool produces wrong output (e.g., fabricated citations, misquotation, erroneous holdings). This requirement pervades the official record: the NYSBA warns that general-purpose models return erroneous results on a large majority of legal queries~\cite[p.20]{nysba2024taskforce} (citing~\cite{dahl2024}); legal educators instruct lawyers to ask if there is ``a known error rate, and if so, is it considered acceptable?''~\cite{cle2026}; and LSOs are advised to ``define baselines and success metrics'' including ``error rates''~\cite[p.29]{nysba2026a2j} (see also~\cite{evidence_advisory_report_2025}).
    
    \item[Report operational metrics.] The system must capture how a tool performs in a workflow (e.g., throughput, rework, the share of output a reviewer accepts unaltered). This is distinct from whether the output is correct, and each metric must be designed carefully: a high first-pass acceptance rate may reflect a reliable tool or a lenient reviewer.
    
    The official record supports this requirement. The \textit{access-to-justice} (A2J) playbook directs organizations to track hours saved, cycle-time reduction, staff satisfaction, and percent of outputs accepted on first review~\cite[p.29]{nysba2026a2j}. It also calls for equitable metrics: ``who benefits (by geography, language, practice area) and who is left out (digital access, disability accommodations)''~\cite[p.29]{nysba2026a2j}.
    
    \item[Support comparative evaluation.] The system must establish whether a tool outperforms the status quo at a defined task. In recent hearing testimony, an LSO general counsel posed seven questions that must be ``answered by the legal services community as a whole''~\cite[p.72]{nysba2026a2j}, including how well genAI can communicate with clients and conduct intake. Lacking comparative legal studies, he drew on data from medicine (e.g.,~\cite{ayers2023}). Lawyers need a domain-specific body of evidence on the costs and benefits of using AI on legal tasks.
    
    \item[Define and measure error constructs.] The system must capture how AI error differs from typical (domain-specific) human error. A representative of pro-bono organizations named two relevant disadvantages of AI error: that the tools are ``confidently wrong'' and ``seen as more reliable'' by humans~\cite[p.92]{nysba2026a2j}. These map to concepts in the research literature. For example, Dahl et al.~\cite{dahl2024} document two relevant behaviors:
    \begin{enumerate}[leftmargin=0.8cm]
        \item \textit{contra-factual bias}, the tendency to accept false premises rather than reject them; and
        \item \textit{miscalibration}, the failure of a model's expressed confidence to track its accuracy.
    \end{enumerate}
    Contra-factual bias is one way to be confidently wrong; miscalibration is one way to over-trust. Both can be measured.
    
    \item[Apply task-conditioned thresholds.] The system must allow users to set reliability thresholds at task granularity. The acceptable rate of fabricated citations in an exploratory research memo differs from that in a filed brief. Users should be able to define their own thresholds (e.g., by task, tool, and input data type), and the system should use them in communicating results.
    
    Threshold determinations themselves are external. The judiciary has drawn one such threshold, holding that (general purpose) genAI tools ``are not suitable for legal writing and legal research''~\cite{ucspolicy2025} (i.e., a ``do-not-use'' determination across an entire task class). Defining thresholds and applying them to results is internal.
\end{description}

\noindent
\textbf{Functional Requirements:  Mechanisms}

These requirements focus on data management and analysis. A \textit{benchmark} can compare AI tools on a standard dataset, but this is not sufficient for legal AI as many legal tasks are highly context-dependent (e.g.,~\cite{withers2026}). The evaluation system must therefore support features that pertain to the data itself.

\begin{description}

    \item[Benchmarks.] The system must provide shared, public test sets paired with reference answers. AI tools can be run against these benchmarks to yield comparable, per-task error figures. The official record calls for this explicitly: legal educators state that ``independent benchmarking'' is critical~\cite{cle2026}, and experts from the \textit{National Center for State Courts} recommend ``independent testing and evaluation for accuracy, reliability, and utility''~\cite[p.61]{nysba2026a2j}.\footnote{See also Franklin et al. \cite{franklin2025evaluating}, who recommend a ``living benchmark that reruns every quarter'' to keep pace with model updates.} A benchmark measures performance on a practical matter only if the matter's task, scale, and document profile fall inside the benchmark's distribution.
    
    Several legal AI benchmarks have been developed (e.g., \cite{ramjee2026redlinebench}). LegalBench~\cite{legalbench2023}, a collaborative effort between the computing and legal communities, provides 162 tasks drawn from 36 data sources, each measuring a specific type of legal reasoning (e.g., issue-spotting, interpretation). But legal reasoning benchmarks are not sufficient to ground the duties described in this paper. Reliance benchmarks must account for failure modes and the cost of establishing ground truth. LegalBench measures whether a model can identify a contract clause type or interpret a statute; reliance requires measuring failure rates on the actual artifacts lawyers produce (e.g., briefs, filings, memos) under realistic conditions.

    \item[Portable test harness.] The system must provide tooling that stakeholders can use to run standard evaluations on their own data. This requirement is implicit in the official record: the A2J playbook asks for ``pilot intake'' studies covering ``use case, data sensitivity, grounding corpus, reviewer, success metrics, review date''~\cite[p.28]{nysba2026a2j}. In most cases, no public benchmark will match the distribution of the actual data.
    
    Practical exemplars exist: general-purpose, open-source evaluation libraries (e.g., the ELM toolset from the Software Engineering Institute~\cite{turri2026elm}) let users gather measurements and metadata (e.g., configuration, model version). Since these tools presuppose technical expertise, reducing the barrier to operation would be a major contribution.

    \item[Ground-truth construction.] The system must provide a method for establishing ground truth on a given task. Without an understanding of correct output, no scores can be generated. The official record calls for ``defining and measuring'' capabilities, but does not suggest how.
    
    Ground truth in legal tasks is rarely uncontested: what counts as correct depends on several factors, including jurisdiction, existing doctrine, and strategic objectives. The system must therefore support multiple ground-truth conventions, from shared reference answers established by practice communities to private oracles defined by individual stakeholders or specialist legal communities (e.g., biotech patent lawyers).

    \item[Conditioning metadata.] The system must support metadata needed to interpret measurements, including task type, input scale, document profile, jurisdiction, and tool version. This requirement is implicit in the official record. Opinion 512 ties verification to the tool and the specific task~\cite{aba_formal_512}, and the NYUCS draws a task-class line~\cite{ucspolicy2025}. The generalization to full metadata is new, motivated by cases (e.g.,~\cite{withers2026}) in which properties of the tool were ignored. Evaluation results should be surfaced with metadata attached.

\end{description}

\subsection{Non-functional Requirements}

The \textit{non-functional requirements} state how the evaluation system should operate. They specify conditions under which measurements remain meaningful and usable.

\begin{figure}[ht]
    \centering
    \includegraphics[width=1\linewidth]{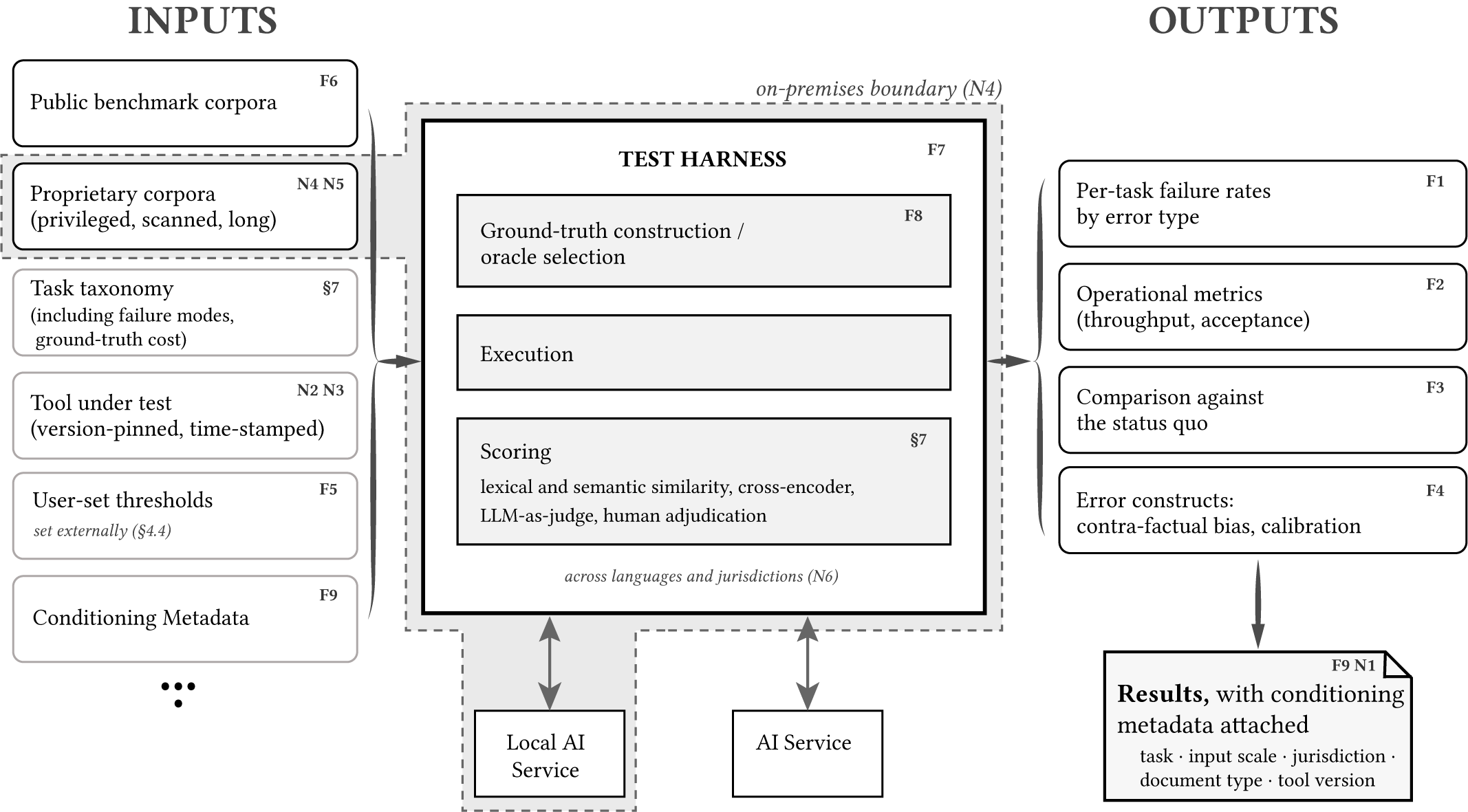}
    \caption{Evaluation Infrastructure for Legal AI.}
    \label{fig3_evaluation_infrastructure}
    \Description{A diagram of the evaluation architecture for legal AI tools, showing inputs, the test harness used to evaluate performance, and outputs.}
\end{figure}

\begin{description}
    
    \item[Task-conditioning (reporting granularity).] The system must support reporting per task, not merely at a global level. Opinion 512 ties verification to ``the GAI tool and the specific task''~\cite{aba_formal_512}, so a single tool-level number cannot discharge it. Per-task measurements (F1) must travel with results (F9) so users understand the context.
    
    \item[Currency.] The system must track tool and data versions. Verification is specific to task and tool, so evidence must be gathered again as models change; the NYCBA deliberately keeps its guidance general because the tools change quickly~\cite{nycbar2024} (see also \cite{Chen2024How}). The law itself moves: a statute may be amended or a case overruled, so an unchanged tool can begin giving stale answers. Input data can also drift, as when a benchmark drawn from a public repository grows stale through error correction.
    
    \item[Provenance.] The system must record lifecycle information for both tools and data. Data provenance can help users understand results or debug unexpected outputs, and knowledge of past data pipelines can be used to reconstruct or enlarge input datasets. Specific legal use cases include ensuring training and evaluation data is properly licensed, and forensic analysis in sanctions proceedings (where the input dataset, tool version, and prompt must be reconstructed).\footnote{\ Prompts are already the subject of rulemaking. For example, the NYUCS retains prompts of judges and court staff as institutional records subject to audit and disclosure~\cite{lebovits2026courts}. At least one New York court has held that a litigant's prompts are conditionally protected as trial preparation material~\cite{assini2026hayward}.}

    \item[Confidentiality.] The system must support strong confidentiality safeguards. Many useful tasks can be explored with public data, but some require sensitive documentation; stakeholders may also wish to use local infrastructure to keep data on premises.
    
    \item[Input robustness.] The system must support the inputs stakeholders actually use, including scanned images, long records, irregular formatting, and complicated tables. A characterization on clean text does not transfer to a 500-page scanned record.
    
    \item[Linguistic and jurisdictional diversity.] The system must support measurements across languages and jurisdictions. LSOs serve clients who speak minority languages; the judiciary already treats language access as a distinct function~\cite{ucspolicy2025}; and measurements made in one jurisdiction or language are generally not commensurable (see~\cite{dahl2024,withers2026}).

\end{description}
Table~\ref{tab:requirements} summarizes the functional and non-functional requirements. Figure \ref{fig3_evaluation_infrastructure} shows a simplified representation of a corresponding evaluation architecture for legal AI tools.

\begin{table*}[t]
\centering
\caption{Requirements for a legal-AI evaluation infrastructure, with the warrant for each.
``R'' = demanded in the NY hearing record or governing instruments; ``E'' =
obtained from measurement and software practice.}
\label{tab:requirements}
\small
\begin{tabular}{p{0.05\textwidth}p{0.30\textwidth}p{0.45\textwidth}p{0.10\textwidth}}
\toprule
& \textbf{Requirement} & \textbf{What the system must do} & \textbf{Warrant} \\
\midrule
\multicolumn{4}{l}{\textit{Functional --- required outputs}} \\
F1 & Report failure rates by task & Per-task rate of wrong output, by measurement & R\\
F2 & Report operational metrics & Workflow measures (e.g., throughput, first-pass acceptance) & R\\
F3 & Support comparative evaluation & Tool vs.\ status quo on a defined task & R \\
F4 & Define \& measure error constructs & Operationalize failure constructs (e.g., contra-factual bias, calibration error) & R \\
F5 & Apply task-conditioned thresholds & The line against which a measured rate is judged & R \\
\multicolumn{4}{l}{\textit{Functional --- required mechanisms}} \\
F6 & Benchmarks & Shared test sets + reference answers; cross-tool & R \& E \\
F7 & Portable test harness & Run the evaluation on the user's own corpus & R \& E \\
F8 & Ground-truth construction & Establish ground truth for an own-data test & E \\
F9 & Conditioning metadata & Validity region per task, scale, profile, jurisdiction, version & E\\
\multicolumn{4}{l}{\textit{Non-functional --- system qualities}} \\
N1 & Task-conditioned reporting & Results carry their task scope; no single global figure & R \\
N2 & Currency & Re-evaluation triggered by changes in tools, law, or evaluation data & R \\
N3 & Provenance & Lifecycle information for tools and data (e.g., for licensing, sanctions reconstruction) & E \\
N4 & Confidentiality & No exposure of privileged/client data under test & R \\
N5 & Input robustness & Scanned, OCR'd, long, irregular inputs & E \\
N6 & Linguistic and jurisdictional diversity & Coverage across languages and jurisdictions & R \\
\bottomrule
\end{tabular}
\end{table*}

\subsection{Adjacent Concerns}

Several concerns are outside the requirements analysis because they depend on external issues rather than properties of the system. First, a low barrier to operation would benefit the legal system as a whole: an LSO should be able to run an evaluation without a data-science team, a licensed corpus, or significant compute.

Second, the evaluation system should also have a means for establishing shared ground truth. Functional requirement F8 supports multiple ground-truth conventions, but does not address how disparate communities arrive at shared ``open oracles,'' who maintains them as the law changes, or how conflicts between competing oracles are resolved. These are questions of institutional authority, not system design.

Third, the requirements state that the evaluation system must support task-conditioned thresholds at any granularity. They do not, however, provide a means of determining those thresholds. Stakeholders will differ greatly in the factors (e.g., risk aversion) that influence threshold choice. This issue is also out of scope.

\newpage
\section{Process as a Substitute for Evidence}
\label{sec:substitution}

Legal AI lacks a suitable body of evidence, and the official record shows that stakeholders are filling the gap with substitutes.
\begin{description}

\item[Policies and procedures.] Many organizations tackle AI governance by policy and procedure. Policy documents typically state that an AI tool may be used if its output is verified, the workflow is documented, and appropriate steps are followed. Policies are convenient because they are auditable; they also produce the illusion of diligence.

But documentation is not enough. A workflow step that says ``verify the output'' assumes the reviewer knows how much verification is enough. In \textit{Withers v. City of Aberdeen}~\cite{withers2026}, a law firm had a written policy requiring independent verification of an AI tool's output; a lawyer filed documents with hallucinations anyway. The policy said to verify, not how much verification this tool and task required (see also \cite{Landberg_City_New_York_2026}). Process cannot fill the gap left by evidence.

\item[Private pilots.] The NYSBA recommends pilot tests with ``clear success metrics and audit trails to build trust and demonstrate value of AI tools''~\cite[p.10]{nysba2026a2j}. A lawyer might test a tool on a handful of documents, judge the results acceptable, and rely on it afterward; Opinion 512 contemplates that prior experience with a tool may warrant less verification~\cite{aba_formal_512}.
Pilot tests are evidence gathering, but unrecorded, unshared, and run on whatever data was at hand. The results are not propagated or reviewed. A pilot is a snapshot in a particular operational context, producing no lasting record.

\item[Human review.] Many organizations require humans to review AI output. Lawyers are asked to conduct a review and certify that it occurred~\cite{nys_courtrules_2026}. Human review catches some errors and misses others: a reviewer can spot a fabricated citation but will struggle to detect a misinterpretation or omission (see Section~\ref{sec:human_review_limits}). Review catches issues but is not a substitute for measurement.\footnote{A sitting New York judge recently considered the same issue, concluding that the only evidence available to courts at the moment is process evidence (e.g., draft histories, metadata, and prompt logs) \cite{lebovits2026courts}.}

\item[Specialists.] Several organizations recommend specialists who can provide information about AI tools. Opinion 512 advises lawyers to draw on others who understand a tool's capabilities and limitations~\cite{aba_formal_512}, and legal educators have urged the same~\cite{nysba_ai_cle_2026_trust_but_verify}. The problem is that there is no settled body of evidence for a specialist to rely upon: the expertise being sought does not exist.

\item[Local measurement.] The official record repeatedly tells lawyers to conduct their own experiments: the A2J playbook directs each provider to establish its own error rates~\cite{nysba2026a2j}, and court-technology experts tell courts to ``look for or conduct'' their own testing~\cite[p.61]{nysba2026a2j}. These measurements are small, done differently in every instance, and incommensurable. The work is handed to the parties least equipped to do it, and redone from scratch everywhere.

\item[Training.] Institutions use training as an administrative control. It is inexpensive, scalable, and useful for demonstrating that issues are being addressed; it is also used for discipline (e.g., the lawyer in \textit{Withers} was ordered to attend training courses on AI tools~\cite{withers2026}).

As with policies, it is easy to mistake prescription for feasibility. Legal educators stress the importance of studying AI tools' capabilities and limitations, but cannot point to a body of evidence. The result is more of the same recommendations: human review, AI policies, specialists, and internal pilots.

\end{description}
Despite their differences, these substitutes share a fatal flaw. The evidence the duties require is shared, enduring, and task-specific; each substitute is the inverse of one of these properties: a private result, a rapidly aging measurement, or a procedure indifferent to context. They are ephemeral proxies. The institutions confronting these tools have not produced the missing evidence, and the substitutes they have adopted do not produce it either.

\section{The Actual Evidence}
\label{sec:evidence}

We have claimed that no body of evidence supports the legal duties analyzed in this paper. The claim must be tested against two apparent counterexamples: (1) a small, but growing, peer-reviewed literature on legal AI tools and; (2) eDiscovery, a well-studied legal task with well-established methods for assessing reliability.

\subsection{Legal AI Research}

Dahl et al.~\cite{dahl2024} provided the first critical evaluation of foundation models for legal reasoning. The results were not encouraging: LLMs produced false information at substantial rates and were poor judges of their own reliability. Their work confirmed that error rates are measurable, but the authors covered a small volume of the $\langle task, tool, dataset \rangle$ space using general-purpose tools. It established that the problem is real, but it does not tell a lawyer how a particular tool will fail on her task with her data.

A later study by Magesh et al. \cite{magesh2025} tested commercial legal research tools. Although some of these tools were supposed to have reduced rates of hallucination, they still produced incorrect answers on a significant percentage of queries. This is closer to practice, but coverage remains narrow: a snapshot of a few products on one set of research questions, with no conditioning on document type or jurisdiction. It is a sample of the evidence that is needed, not the evidence itself.

Salinas et al.~\cite{salinas2026} examined AI tools in the context of contract law. Law professors compared AI answers to answers from other law professors, showing a general preference for AI. The study did not measure correctness against a known standard; instead, success  was measured by expert judgment on which answer was better for students. This approach suits the study's educational purpose but is not a good fit for an evaluation of reliability.

Liu et al. \cite{liu2026} measured the accuracy of AI models on citation verification, using a benchmark  of legal briefs with injected errors. No system combined high recall with high precision: the highest-recall detector found 83\% of errors but flagged more correct citations than incorrect ones. Errors were not detected equally: (1) \textit{self-announcing failures} like non-existent cases and mismatched names were found reliably; (2) incorrect pincites and content misrepresentations were not. Hallucination rates did not decline monotonically across model generations (i.e., newer models are not necessarily better).

These studies are useful exemplars. Empirical evaluations of legal AI appear to be produced at a slow rate, while the field itself is changing rapidly. The studies that we surveyed do not measure the same tasks, as they cover: (1) hallucinations in general-purpose models; (2) errors in commercial research tools; (3) preferences over AI and human answers in an educational context; and (4) citation verification. Each study is a snapshot of particular tools as they are applied to specific tasks. A handful of academic works cannot provide a body of evidence that allows a lawyer to assess the capabilities and limitations of an AI tool on the matter in front of her.


\subsection{eDiscovery and TAR}

A robust body of evidence does exist for one narrow area: eDiscovery. TAR uses software to find relevant information in datasets, a form of information retrieval whose reliability can be measured via statistics on labeled samples. Courts have accepted TAR on that basis for over a decade~\cite{grossman_cormack_2011,da_silva_moore_2012}. This is exactly the characterized, defensible evidence missing everywhere else in legal AI.

TAR has three properties that make the measurement problem tractable: (1) the task is classification over a fixed set of documents; (2) the correct answer for a sample can be established by human review at reasonable cost; and (3) the relevant standard is itself a rate, so a measured rate is an appropriate answer. Together, these give rise to a body of knowledge that can be shared across the field. These are precisely the properties genAI evaluation lacks.

The field is also an example of cross-discipline collaboration. The infrastructure and methodology originated from outside the legal profession: computing researchers devised the evaluation methods and built the software to perform measurements, while the legal system provided domain knowledge and feedback. The body of evidence was not generated by lawyers piloting tools in their offices. It was generated by collaboration between the legal and computing communities.

\subsection{Toward a Body of Evidence for genAI}

No body of evidence supports the legal duties, with the narrow exception of eDiscovery. That example suggests that when the task has the right properties, and legal experts are supported by the computing community, legal AI can be characterized to a standard that courts will accept.

Unfortunately, GenAI tools differ from TAR in ways that defeat each of the three properties above. First, human review of some genAI outputs requires as much work as producing the output anew. Second, a generated document is unique, precluding the standard statistical estimates of reliability that depend on repeated trials with stable ground truth. Third, the relevant standard is not always a rate: a single fabricated citation in a filed brief is a different kind of failure than a 5\% miss rate in document review.

Measurements of genAI tools occupy a small region in a large space of conditions (F9). Building a comparable body of evidence will require sustained, cross-discipline work of the kind that produced TAR, applied to a problem with fewer of the conveniences that made TAR tractable.

\section{A Research Agenda}
\label{sec:agenda}

This section provides a brief research agenda for meeting the requirements defined in Section \ref{sec:requirements}. Due to space constraints, we cannot present a complete agenda; instead, the goal is to list important open problems. As with the requirements, the focus in this section is on building the evaluation system. Governance and ethics are out of scope. 
\begin{description}
    \item[Task taxonomies.] Use cases should be decomposed into tasks, and tasks should be organized into taxonomies. A usable taxonomy should: (1) split legal work into fine-grained tasks that allow for informative measurements (e.g., error rate); and (2) distinguish these tasks by their data requirements (e.g., whether a public benchmark suffices, whether private testing is required, or whether correctness is contested).
    
    Existing legal task taxonomies categorize tasks by reasoning type or cognitive level. For instance, LegalBench ~\cite{legalbench2023} groups tasks by legal reasoning (e.g., issue spotting, rule recall), while other works group tasks by cognitive level (e.g.,~\cite{legaleval2024}). These taxonomies are useful for distinguishing tasks based on \textit{capability}, not \textit{reliability}. The legal duties require taxonomies that also account for failure modes (e.g., fabrication, omission, misinterpretation) and the cost of establishing ground truth. The evaluation system should support multiple competing taxonomies, letting users select those that meet their needs.
    
    \item[Establishing reference answers.] Every test needs a standard to score against. For genAI tools, it is often difficult to determine whether a given output is acceptable (e.g., whether a legal argument is correct, or whether two case summaries are equivalent). The central research problem is to lower the cost of construction: methods for building reliable reference answers without redoing the task by hand every time. Such methods would allow greater participation from a legal community already struggling to meet service demand.
    
    \item[Scoring outputs.] A set of reference answers does not by itself make scoring automatic. The evaluation system must judge how closely a tool's output corresponds to the reference. For TAR this is straightforward; for open-ended legal tasks it is an active research problem. 
    
    The system needs a validated, task-dependent catalog of legal similarity measures. Traditional measures are not sufficient: lexical-overlap measures favor shared wording, which can be misleading on legal tasks, and semantic-similarity measures capture topical distance rather than legal argument. More sophisticated methods include cross-encoder similarity and ``LLM-as-judge,'' both of which require a second AI tool that must itself be evaluated. The most reliable (and least scalable) method is human judgment.
    
    \item[Benchmarks.] Requirement F6 calls for shared benchmarks. The research problem is sustaining them: existing legal benchmarks~\cite{legalbench2023,legaleval2024} measure what tools can do, not how often they fail, and most ML benchmarks are not maintained after release~\cite{betterbench2024}. For tools revised every few months, an unmaintained benchmark cannot calibrate reliance, and for most legal work (unlike narrow, document-uniform fields such as patents or tax filings) no public benchmark will reflect the distribution of actual matters (see~\cite[p.231]{magesh2025}).
    
    \item[Test harnesses.] Requirement F7 calls for tooling that allows users to evaluate AI tools on their own corpus. The research problem is reducing the barrier to operation so that evaluation is available to LSOs and other organizations that lack computing expertise.
    
    \item[Conditioning results.] Requirement F9 calls for metadata that travels with results. The research problem is determining which conditions actually change a tool's reliability: where are the boundaries of a result, and how does output quality change as those boundaries are tested? A positive result in one setting can be misread as a license to rely on the tool in others. Better guidance  would help lawyers assess whether a tool applies to a given matter, and help courts determine whether reliance was justified.
\end{description}

\subsection{Challenges}

Three challenges arise for the agenda above. First, commercial AI tools are largely \textit{black boxes}: users can only observe their outputs. Recent research has shown that black-box access alone is insufficient for rigorous auditing, since some failure modes require examining the model's internal behavior~\cite{casper2024}. The evaluation system can only confirm failures visible at the output, so its results overstate reliability. Measurements under black-box conditions are therefore lower bounds on the actual failure rate.

Second, AI paradigms are evolving rapidly. This article has focused on chat-based tools, but an emerging agentic ecosystem (with stateful interactions, retrieval from external datastores, complex orchestration, and tool-to-tool handoffs) multiplies the dimensions along which results must be conditioned. Versioning and provenance become harder, and some agentic failure modes may have no analog in the single-call setting the agenda has assumed.

Third, the duties analyzed here were extracted from one state and a small amount of federal documentation. Other US states have issued guidance with different emphases (e.g., \cite{ohnd2023boyko,ilsupct2025aipolicy}), and other common-law and civil-law jurisdictions are active in this space. A comparative survey would either strengthen the analysis by showing convergence in legal duties, or qualify it by revealing that the duties presented here are jurisdiction-specific.

\section{Conclusion}
\label{sec:conclusion}

This paper synthesized the legal duties governing the use of AI in the New York court system. A lawyer must (1) understand a tool's capabilities and limitations; (2) verify its output in proportion to task and tool; and (3) review and certify before filing. The review and certification duties apply equally to lawyers and SRLs.

Together, the duties amount to a demand for \textit{appropriate reliance}: relying on a tool when evidence supports it, declining or verifying when it does not. The HCI literature has analyzed appropriate reliance at length, but its application here turns on a body of evidence the legal system has not produced. The first duty presupposes evidence about what tools can and cannot do; the second presupposes methods for verifying outputs that may be costly or infeasible to verify; the third makes the lawyer the backstop when both fail. All three presuppose evidence that does not exist for the genAI tools proliferating across the legal system.

This paper made four contributions: (1) it mapped the legal duties onto concepts from the HCI literature on appropriate reliance; (2) it elicited a set of requirements for an evaluation infrastructure from the official record; (3) it documented the procedural substitutes the legal system has adopted in place of evidence (e.g., policies, human review); and (4) it sketched a research agenda for the computing community.

Collaboration between the computing and legal communities produced the one area of legal technology backed by a sufficient body of evidence: eDiscovery. The reliability of TAR tools is measured with statistical methods, validated against samples, and accepted by courts. That evidence was not produced by lawyers piloting tools in their offices; it was produced by sustained collaboration.

\newpage
The same collaboration is needed again. The official record shows that courts and bar associations recognize the risks of legal AI but have little evidence to draw on. The computing community has the methodological resources to build that evidence; the legal community has the domain knowledge to define what must be measured. Until the work is done, the duties remain in force but the means of discharging them do not, a burden that falls hardest on those who are least able to bear it.

\bibliographystyle{ACM-Reference-Format}
\bibliography{references}

\end{document}